\documentclass{emulateapj}

\begin{document}

\title{Probing the Magnetic Field Structure in Gamma-Ray Bursts\\
through Dispersive Plasma Effects on the Afterglow Polarization}

\author{Amir Sagiv\altaffilmark{1},
Eli Waxman\altaffilmark{1} and Abraham Loeb\altaffilmark{2,3}}
\altaffiltext{1}{Physics Faculty, Weizmann Institute of Science,
Rehovot, Israel; amir@wicc.weizmann.ac.il}
\altaffiltext{2}{Astronomy Department, Harvard University, 60
Garden St., Cambridge, MA 02138, USA;} \altaffiltext{3}{Einstein
Minerva Center, Physics Faculty, Weizmann Institute of Science}


\begin{abstract}

The origin and structure of magnetic fields in Gamma-Ray Burst
(GRB) fireball plasmas are two of the most important open
questions in all GRB models. Recent measurements of $\gamma-$ray
polarization were claimed to suggest the presence of a uniform
field originating in the compact object driving the outflow. This
interpretation is, however, controversial, since a high degree of
linear polarization is also possible in the presence of a random
magnetic field, arguably originating in electromagnetic
instabilities that develop at the collisionless shock. We show
that the structure and strength of the magnetic field may be
constrained by radio and IR observations of the early afterglow,
where plasma effects on the polarization of the propagating
radiation are significant. We calculate these propagation effects
for cold and relativistic plasmas, and find that in the presence
of a uniform equipartition field the degree of linear polarization
is suppressed and circular polarization prevails at low
frequencies, below $\sim 1-3$ GHz ($\sim \textrm{few} \times
10^{14}$ Hz) in the forward (reverse) shock, at the onset of
fireball deceleration. At higher frequencies linear polarization
dominates. At the frequency of the transition between circular and
linear polarization, the net level of polarization is minimal,
$\sim 10$--20\%. These features are nearly independent of the
density of the environment into which the fireball expands. When
the uniform field is much weaker than the equipartition value, the
transition frequency is smaller by an order of magnitude.
Depending on the geometry of the emitting region, oscillations of
the polarization position angle may be observed from the optical
reverse shock emission, provided that the strength of the magnetic
field is close to equipartition. The dependence of these results
on viewing geometry, outflow collimation and magnetic field
orientation is discussed. When the field is entangled over length
scales much smaller than the extent of the emitting plasma, the
aforementioned effects should not be observed, and a linear
polarization at the few \% level is expected. Polarimetric
observations during the early afterglow, and particularly of the
reverse shock emission, may therefore place strong constraints on
the structure and strength of the magnetic field within the
fireball plasma.

\end{abstract}


\keywords{
gamma rays: bursts --- magnetic fields --- plasmas ---
polarization --- shock waves}


\section{Introduction}
\label{sec:Introduction}

The past decade has been marked by major advances in our understanding of
Gamma-Ray Bursts (GRBs). These bright explosions at cosmological distances,
are now commonly believed to be generated by the dissipation of a
relativistic outflow associated with the gravitational collapse of a
stellar mass object to a compact remnant (for reviews, see Piran 2000;
M\'esz\'aros 2002; Waxman 2003a). Observations suggest that the outflow
is probably collimated to within a narrow jet, with smaller opening angles
corresponding to apparently more energetic bursts (Frail et
al. 2001). Current models relate the production of $\gamma-$rays to
emission of synchrotron or inverse-Compton radiation by relativistic
electrons. Subsequently, the interaction of the ejecta with the circumburst
gas drives a forward shock into the ambient medium, and a reverse shock
into the ejecta. This stage during which the forward and reverse shocks
coexist is usually dubbed the ``transition phase''; once the reverse shock
fully crosses the ejecta, the dynamics of the expanding fireball is
adequately described by a self­similar solution (Blandford \& McKee
1976). During the self-similar phase which follows, a forward shock
propagates in the ambient gas, gradually decelerating, until the inverse of
the shock Lorentz factor drops below the jet opening angle (indicating the
``jet break time''), at which stage the jet quickly slows down and spreads
sideways.  Subsequently the expansion continues sub-relativistically.
Electrons accelerated at these shocks emit synchrotron radiation, which
appears as an optical flash from the reverse shock (M\'esz\'aros, Rees
\& Papathanassiou 1994; Akerlof et al. 1999; Zhang, Kobayashi \&
M\'esz\'aros 2003, and reference therein),
lasting typically for the duration of the GRB itself,
as well as a long-lived transient afterglow from the forward shock,
characterized by a decreasing peak frequency, which lasts for weeks or even
months (Frail, Waxman \& Kulkarni 2000).

Despite the successes of the models in predicting and explaining light
curves and fluxes at various epochs and various wave­bands, key questions,
concerning predominantly the nature of the progenitor, the launching of the
relativistic outflow, and the physics of the relativistic shock waves, are
still unanswered.  In addition, the physical mechanisms responsible for the
production of $\gamma-$rays are still subject to controversy.  Leading scenarios for
the production of the prompt $\gamma-$ray emission can be grouped into
three main categories. The first consists of models in which the
relativistic outflow is dominated by the kinetic energy of the ejecta, and
the $\gamma-$rays are the consequence of synchrotron emission by
relativistic electrons, which are Fermi-accelerated in optically-thin
collisionless shocks created by inhomogeneities in the ejecta (``internal
shocks''). The assumption that the $\gamma-$rays are synchrotron photons
requires strong magnetic fields, which probably result from the
amplification of a small scale field by plasma instabilities in
collisionless shocks (Gruzinov \& Waxman 1999; Medvedev \& Loeb 1999).
In this scenario, the coherence scale of the magnetic field may grow
to the size of the causally connected regions in the outflow, a size
which is still smaller than the scale of the emitting slab.
Afterglow observations strongly favor the formation of strong
magnetic fields at relativistic shock fronts, and may therefore
indicate that such a mechanism is also governing the emission from
the internal shocks.
In a second type of models the outflow
is Poynting-flux dominated, and current-driven instabilities at large radii
lead to acceleration of pairs that emit $\gamma-$rays by synchrotron
radiation (Lyutikov, Pariev \& Blandford 2003).
The third type of models associate the
prompt emission with the relativistic boosting of an ambient photon field,
possibly related to an earlier SN explosion (Lazzati et al. 2000), by
electrons outflowing with a high Lorentz factor.

The recent detection of a strong linear polarization signal in the prompt
$\gamma-$ray emission of GRB 021206 by Coburn \& Boggs (2003) triggered
much interest in GRB polarization features [but see Rutledge \& Fox (2003),
who critically reanalyze the lightcurve of GRB 021206, and find no
polarization signal]. This measurement -- if real -- provided a strong
indication that synchrotron emission is indeed the underlying radiation
mechanism of the GRB phenomenon (Waxman 2003b). Its interpretation in terms
of magnetic field structure is, however, ambiguous. The high polarization
in $\gamma$-rays was claimed to suggest the presence of a uniform magnetic
field in the emitting plasma (Lyutikov et al. 2003), thus supporting the
paradigm of a Poynting-flux dominated outflow advecting a ``primordial''
field from the ``inner engine''. It turns out, however, that this line of
reasoning is not free of caveats. First, it was recently demonstrated
(Waxman 2003b; Granot 2003; Nakar, Piran \& Waxman 2003) that an outflow
which is collimated to within a narrow jet and is observed off axis,
may exhibit a high degree of linear polarization, even in the presence
of a random magnetic field.
Second, even the presence of a uniform field is still not
an indication that the outflow is Poynting-flux dominated, since the
amplitude of a toroidal magnetic field decreases as $R^{-1}$ and tends to
dominate at large radii -- even if the outflow is dominated by kinetic
energy.  The uniform field hypothesis is controversial also since it
appears difficult to reconcile the existence of a large scale magnetic field with the
presence of relativistic charge particles, which are accelerated in situ,
since current understanding of particle acceleration requires the dissipation
of these large-scale fields.

In this paper we examine the fireball scenario, in which the outflow is
dominated by the plasma kinetic energy, and demonstrate that the strength
and structure of the magnetic field may be constrained by radio and IR
observations of the early afterglow, where plasma effects on the
propagation of synchrotron radiation are significant. Specifically, we show
that the presence of a uniform field permeating the plasma has distinctive
fingerprints on the polarization properties of emitted radiation, which
discriminate it from the case of a field entangled over small length
scales. Of particular interest is the effect on the polarization of the
reverse shock emission, since that plasma is just the ejecta producing the
prompt $\gamma-$ray emission. Hence constraining the structure and strength
of the magnetic field in the reverse shock is complementary to measurements
of the $\gamma-$ray polarization, and has direct implications for the plasma
conditions in the ejecta and consequently for processes taking place in the
vicinity of the compact source. This powerful probe may help to put
constraints on models of field generation, thus promoting our understanding
of both the shock physics and the nature of the ``inner engine''.

This article is organized as follows.
In \S\,\ref{sec:Propagation_Effects} we provide a brief review of
polarized light transfer in a magnetoactive plasma, along with new
results for a relativistic plasma. Next
(\S\,\ref{sec:Field_Direction_and_Viewing_Geometry}) follows a
qualitative discussion of fireball geometry and observation geometry,
in the context of plasma propagation effects.
The propagation effects are applied to GRB plasmas in
\S\,\ref{sec:Application_to_GRB_AG}. In \S\,\ref{sec:Plasma_Parameters} we
obtain plasma conditions during the early afterglow, considering both cases
of an expansion into a uniform-density interstellar medium (ISM) in
\S\,\ref{sec:Plasma_Parameters_ISM} and an expansion into a circumburst
wind in \S\,\ref{sec:Plasma_Parameters_Wind}.  Detailed analysis of
propagation effects at the corresponding shocks follows in
\S\,\ref{sec:Observational_Consequences_for_Uniform_Strong_Field}.  The
implications of the strength and structure of the magnetic field are
discussed in \S\,\ref{sec:Magnetic_Field_Strength_and_Structure}.
Finally, we consider the implications of our
results in \S\,\ref{sec:Conclusion}.



\section{Plasma effects on the polarization of propagating
synchrotron radiation} \label{sec:Propagation_Effects}

As is well known, a plasma permeated by a magnetic field is birefringent,
and a phase shift accumulates between the normal modes as radiation
propagates through the plasma. The observational outcome of this phenomenon
is the Faraday rotation effect.  Below we briefly review the formalism of
transfer of polarized synchrotron radiation in a magnetoactive plasma
(\S\,\ref{sec:Brief_Formalism}), and bring new results for a plasma
dominated by a relativistic electron population
(\S\,\ref{sec:New_Results_for_Relativistic_Plasma}). The appendix gives a
more comprehensive derivation of these results.

\subsection{Formalism}
\label{sec:Brief_Formalism}

The effect of the magnetoactive plasma on the transfer of
polarized radiation may be treated most straightforwardly via the
analysis of the wave equation, incorporating the effect of the
medium by a dielectric tensor. When the radiation can be regarded
as transverse, the tensor $\kappa_{ij} \;\; (i,j =
1,2)$~\footnote{We adopt a right-handed system of coordinates with the
wave-vector $\mathbf{k}$ directed along axis 3.} representing
the plasma effect in the dielectric tensor, is separated into a
sum of hermitian and anti-hermitian tensors as follows~:
\begin{equation}\label{eq:susceptibility_tensor_in_intro_section}
4 \pi \frac{\omega}{c} \kappa_{ij} = \left( \begin{array}{cc}
h & i f \\
-i f & -h
\end{array} \right)
+ i \left( \begin{array}{cc}
\kappa+q & i v \\
-i v & \kappa-q
\end{array} \right)\;.
\end{equation}
The anti-hermitian component of the tensor is responsible for
absorption. The hermitian part is responsible for propagation effects;
specifically, $f=\Delta k(\omega) = \omega \Delta n(\omega) / c$ is the
difference between the wave numbers of the normal modes, giving rise to the
Faraday rotation effect. In the case of synchrotron self-absorption,
the parameter $v$ is typically smaller than $\kappa$ by a factor $\gamma_m$,
where $\gamma_m$ is a typical Lorentz factor of the relativistic electrons,
and will therefore be neglected hereafter.

The wave equation for the Fourier components of the electric field
can be manipulated, with the help of equation
(\ref{eq:susceptibility_tensor_in_intro_section}), into a transfer
equation for the Stokes parameters~:
\begin{equation}\label{eq:Transfer_Equation_for_Stokes_Parameters}
\left\{ \begin{array}{lll} \textrm{d}I / \textrm{d}s &=&
\varepsilon_I - \kappa I - q Q
\nonumber\\
\textrm{d}Q / \textrm{d}s &=& \varepsilon_Q - q I - \kappa Q
- f U \nonumber\\
\textrm{d}U / \textrm{d}s &=& f Q - \kappa U - h V \nonumber\\
\textrm{d}V / \textrm{d}s &=& \varepsilon_V + h U - \kappa V\;,
\end{array} \right.
\end{equation}
where $s$ is a length parameter along the path of the ray, and
$\varepsilon_I, \varepsilon_Q$ and $\varepsilon_V$ are the
emission coefficients corresponding to $I, Q$ and $V$.

The physical significance of the propagation parameters $f$ and
$h$ is most easily demonstrated in a transparent medium. As an
example, consider the case $f \gg h$, with negligible $\kappa$ and
$q$. Equation (\ref{eq:susceptibility_tensor_in_intro_section})
then implies that the normal modes are (left- and right- )
circularly polarized. Integrating equation
(\ref{eq:Transfer_Equation_for_Stokes_Parameters}) we obtain
oscillations in the Stokes parameters $Q$ and $U$, leading to
oscillations and damping in the degree of emergent linear
polarization~:
\begin{equation}\label{eq:Pi_L}
\Pi_L = \frac{\sqrt{Q^2+U^2}}{I} \simeq
\frac{\epsilon_Q}{\epsilon_I} \left| \frac{\sin(fs/2)}{fs/2}
\right|\;.
\end{equation}
A large phase shift accumulated between the normal modes, $\Delta\phi = fs
\gg 1$, results in damping of the degree of linear polarization by a factor
of $\Delta\phi$. This is accompanied by oscillations of the polarization
position angle $\chi = \onehalf \tan^{-1} (U/Q)$, providing the Faraday
rotation effect.  Analogous, simple to interpret phenomena characterize
also the case $h \gg f$ (linearly polarized normal modes).

\subsection{New results for relativistic plasma}
\label{sec:New_Results_for_Relativistic_Plasma}

The coefficients $f$ and $h$ are easily derived when the plasma is
cold~:
\begin{eqnarray}\label{eq:f_and_h_cold}
f_{cold} &=& \frac{\widetilde{\omega}_p^2 \widetilde{\omega}_B
\cos \vartheta}
{c (\omega^2-\widetilde{\omega}_B^2)} \nonumber\\[3 mm]
h_{cold} &=& \frac{\widetilde{\omega}_p^2 \widetilde{\omega}_B^2
\sin^2 \vartheta} {2 c \omega(\omega^2-\widetilde{\omega}_B^2)}\;,
\end{eqnarray}
where $\widetilde{\omega}_p^2 = 4\pi n_e e^2/m_e$ is the
non-relativistic electron plasma frequency, and
$\widetilde{\omega}_B = e B / m_e c$ is the non-relativistic
electron Larmor frequency. \,$\vartheta$ is the angle between
$\mathbf{k}$ and $\mathbf{B}$, measured in the frame comoving with
the plasma.

In the context of GRB plasmas, however, one also encounters a situation in
which the plasma is dominated by a relativistic electron population. In our
subsequent derivation of analytic expressions for the propagation
coefficients in a relativistic plasma we assume that~: \newcounter{foo}
\stepcounter{foo}({\it\roman{foo}}) the electrons are highly relativistic,
characterized by a spectrum with a power-law tail extending to high
energies; \stepcounter{foo}({\it\roman{foo}}) their distribution function
is isotropic; and \stepcounter{foo}({\it\roman{foo}}) the deviation of the
dielectric tensor from its vacuum value is small. The first assumption is
implied by afterglow observations, whereas the second is adopted for
simplicity. The third assumption will be justified later for the specific
parameters of GRB plasmas. In addition, we restrict the discussion to a
frequency range $\omega \gg \omega_B$, so that modifications to synchrotron
emission and self-absorption coefficients due to the discrete nature of low
harmonics of $\nu_B$ could be safely ignored.

Our main results are summarized in the following equations (for an
outline of the derivation, see
Appendix \ref{sec:Propagation_Coefficients_Appendix})~:
\begin{equation}\label{eq:f_rel_delta_1}
f_{rel}(\nu) \simeq \frac{\widetilde{\omega}_p^2 \widetilde{\omega}_B
\cos \vartheta}
{c \omega^2}\frac{\ln \gamma_m}{\gamma_m^2}
=\frac{e^3 n_e B \cos \vartheta}{\pi m_e^2 c^2
\nu^2} \frac{\ln \gamma_m}{\gamma_m^2}\;,
\end{equation}

\begin{eqnarray}\label{eq:h_rel}
h_{rel} &\simeq& \frac{e^4 n_e B^2 \sin^2 \vartheta}{4 \pi^2 m_e^3
c^3 \nu^3} \nonumber\\
&\times& \left\{ \begin{array}{lll}
\displaystyle \frac{\gamma_m}{2} \left( \frac{\nu}{\nu_m}
\right)^{4/3}
& \;\textrm{if}\; & \nu \ll \nu_m\\[4 mm]
\displaystyle \left( \frac{2}{p-2} \right) \gamma_m^{-(p-2)}
\left( \frac{\nu}{\nu_m} \right)^{(p-2)/2}
& \;\textrm{if}\; & \nu > \nu_m.\\[4 mm]
\end{array} \right.\nonumber\\
&&
\end{eqnarray}
Here $\nu=\omega/2\pi$, $\gamma_m$ is a characteristic electron Lorentz
factor (typically that of the thermal electrons), $\nu_m =
(3eB)/(4\pi m_e c) \gamma_m^2$ is the characteristic synchrotron
frequency corresponding to $\gamma_m$, and $p$ is the spectral index
of the power-law tail of the electron distribution function.
Equation (\ref{eq:h_rel}) generalizes an earlier result by Sazonov
(1969), who considered only the frequency range $\nu \gg \nu_m$.

After the completion of this work, we learned of an independent
derivation of similar expressions for the propagation coefficients
in the relativistic regime by Matsumiya and Ioka (2003).
Our expressions [eqs. (\ref{eq:f_rel_delta_1}) and (\ref{eq:h_rel})]
agree nicely with theirs; nevertheless the resulting polarization
differs substantially, as we further discuss in \S\,\ref{sec:Conclusion}.

Using these results in the context of Faraday rotation, we notice
that $f$, and therefore also $\Delta\phi$, are proportional to
$\nu^{-2}$. The frequency interval $\Delta\nu$, corresponding to a
single period of the position angle oscillation, translates to an
increase of $2\pi$ in $\Delta\phi$, from which we get (for a cold
plasma)~:
\begin{equation}\label{eq:Delta_nu_over_nu}
\left| \frac{\Delta\nu}{\nu} \right| = \frac{\pi^2 m_e^2 c^2
\nu^2}{e^3 n_e B W}\;,
\end{equation}
where $W$ is the width of the slab through which the radiation
propagates. The generalization for a relativistic plasma is trivial.


\section{Consequences of fireball geometry and viewing geometry}
\label{sec:Field_Direction_and_Viewing_Geometry}

A strong polarization signal clearly requires a departure from
spherical symmetry, either via a non-spherical outflow
configuration and a special location of the observer with respect
to it, or through a preferred direction of the magnetic field in
the plasma (or, of course, a combination of the two).
Geometrical considerations related to the fireball collimation,
viewing angle and field's direction are essential for the calculations
we describe below, and so before presenting detailed results of these
calculations, we make a short digression, and discuss qualitatively
some important consequences of the geometry. In particular, we shall
show that the angle $\vartheta$, on which the propagation coefficients
$f$ and $h$ depend [c.f. equations (\ref{eq:f_and_h_cold}) -
(\ref{eq:h_rel})] is not likely to assume special values such as $0$
and $\pi/2$ in a typical GRB jet, and may approach 0 only for a
narrow jet observed off-axis, with a magnetic field oriented normal to
the shock plane; These two regimes have markedly different observational
consequences. We focus in this section only on the case of a large scale,
uniform magnetic field. The case of a fluctuating field is treated in
\S\ref{sec:Magnetic_Field_Strength_and_Structure}.

\begin{figure}[h!]
\epsscale{1}
\plotone{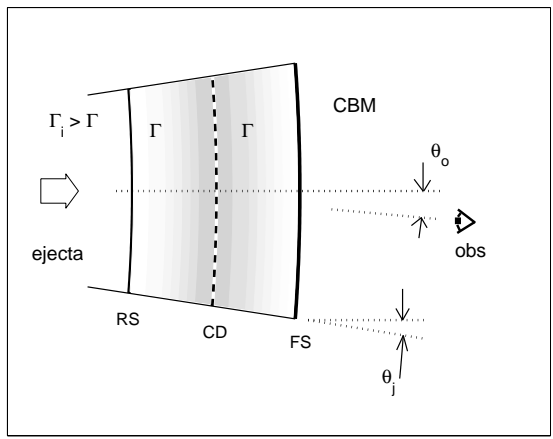}
\caption{Schematic illustration of the fireball and viewing geometry.
$\theta_j$ and $\theta_o$ are, respectively, the jet opening angle and
the angle between the line of sight to the observer and the jet axis.
$\Gamma_i$ is the Lorentz factor of the ejecta prior to deceleration,
whereas $\Gamma$ is the Lorentz factor of the plasma shocked by the
reverse and forward shocks. The initials \emph{RS}, \emph{CD} and
\emph{FS} denote the reverse shock, the contact discontinuity and the
forward shock, respectively, while \emph{CBM} denotes the circumburst
material.
The dark shading marks the cooling of the electrons, with darker shade
representing colder electrons.
\label{fig:FireballGeometry}}
\end{figure}

Fig. \ref{fig:FireballGeometry} shows schematically the essential
ingredients of the fireball and viewing geometry. The figure also
illustrates (with \emph{shaded} areas) electron cooling due to
synchrotron losses away from the (forward and reverse) shock fronts.
This effect, which has important consequences for the resulting
polarization, is discussed in greater detail in
\S\,\ref{sec:Observational_Consequences_for_Uniform_Strong_Field}.

Mounting evidence strongly suggests that the relativistic outflow in GRBs
is collimated into a jet, the opening angle of which is usually inferred
based on estimates of the jet break time from optical observations (Frail
et al. 2001).  These observations are carried on a day timescale after the
GRB trigger, when the jet has already decelerated to a low Lorentz factor,
$\Gamma \sim 10$, whence one deduces $\theta_j \sim 0.1$.  We, on the other
hand, are interested in the emission from the early afterglow, at which
stage the fireball expands with a much higher Lorentz factor, $\Gamma \sim
10^{2.5}$ [see eqs. (\ref{eq:Gamma_FR_ISM}) and
(\ref{eq:Gamma_FR_Wind})]. It is unclear whether during this early stage
the jet opening angle is similar to the one inferred from optical
observations. However, Freedman \& Waxman (2001) have demonstrated that the
isotropic-equivalent fireball energy may be robustly estimated from the
X-ray afterglow flux on a day timescale after the GRB, and furthermore that
this estimate exhibits a strong correlation with the isotropic-equivalent
energy inferred from the GRB $\gamma$-ray fluence [see also Berger,
Kulkarni \& Frail (2003) for a recent analysis compiling more X-ray data].
This correlation, and in particular the fact that the latter estimate is
never much larger than the former, strongly argue in favor of a similar jet
opening angle during the GRB and early afterglow phases. Since the X-ray
afterglow is obtained on a day timescale, when the jet has already
decelerated to $\Gamma \sim 10$, this arguments suggests an opening angle
$\theta_j \sim 0.1$ for a typical GRB jet. An observer of the early
afterglow is therefore not likely to see the edge of a typical jet, and is
consequently not sensitive to deviations of the fireball from spherical
symmetry (other than a large-scale magnetic field in a particular
direction).

Deviations from spherical symmetry may be manifested for an observer
located on a line of sight which makes an angle $\sim \Gamma^{-1}$
from the jet edge, and are therefore probable only if the jet is
extremely narrow, with $\theta_j \sim \Gamma^{-1}$. During
the GRB and early afterglow phases the fireball expands with
$\Gamma \sim 10^{2.5}$, and so the arguments presented above suggest that
such narrow jets are not generic; However, in the framework of the
``standard energy reservoir'' hypothesis promoted by Frail et al.
(2001), brighter GRBs are naturally associated with narrower jets.
Adopting this energy-angle relation, GRB 021206, which was among
the strongest bursts ever, had an opening angle $< 1 / 40$ rad,
or even smaller (Nakar et al. 2003). This observation is key in
associating the high degree of linear polarization measured in GRB
021206 with an observational alignment effect (Waxman 2003b,
Nakar et al. 2003).

\begin{figure}[h!]
\epsscale{1.15}
\plottwo{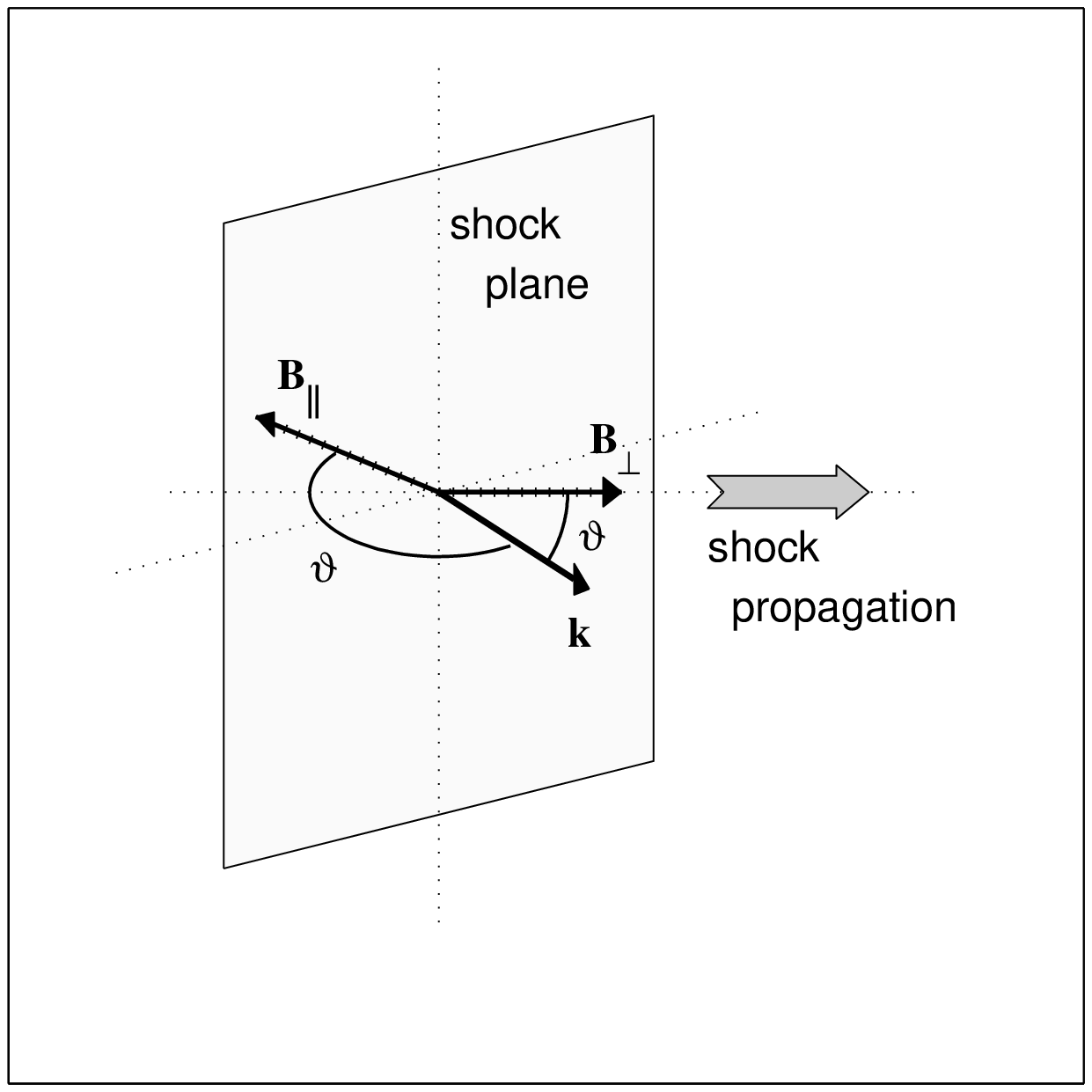}{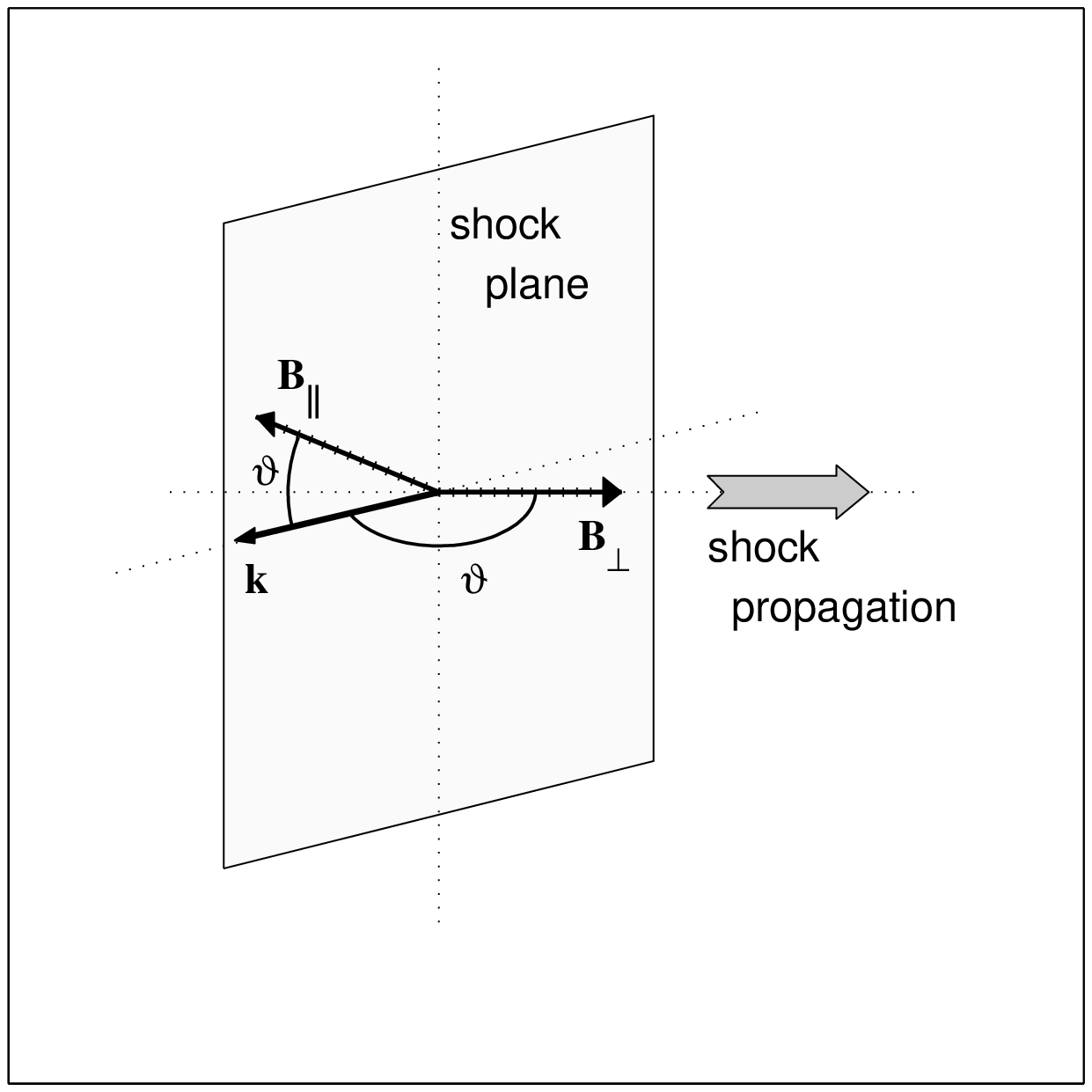}
\caption{Geometrical setup in the frame comoving with the plasma.
Shown are $\mathbf{k}$ --- the direction of the radiation wave vector
(i.e. the line of sight), ${\mathbf B}_{\perp}$ --- a component of the
magnetic field normal to the shock plane, and ${\mathbf B}_{\|}$
--- the field component lying in the shock plane. The angle
$\vartheta$ between the line of sight and the field (either
component) is also shown.
\emph{Left}~: A typical viewing geometry for the case $\theta_j
\gg \Gamma^{-1}$; The wave vector $\mathbf{k}$ lies outside the
plane of the shock. \emph{Right}~: The viewing geometry for the
case $\theta_j \sim \Gamma^{-1}$; Now $\mathbf{k}$ lies roughly
in the shock plane.\label{fig:Geometry}}
\end{figure}

Upon transforming to the frame comoving with the fireball, the
distinct geometries of a \emph{typical jet} and a \emph{narrow jet}
correspond to the radiation $\mathbf{k}$ vector making either a
large ($\gg \Gamma^{-1}$) or a small ($\sim \Gamma^{-1}$) angle
relative to the plane of the shock, for a typical ($\theta_j \gg \Gamma^{-1}$)
and narrow ($\theta_j \sim \Gamma^{-1}$) jets, respectively.
Assume now that the magnetic field is dominated by a component
either parallel or normal to the plane of the shock. The angle
$\vartheta$ between $\mathbf{k}$ and $\mathbf{B}$ may assume
special values, i.e. close to $0$ or to $\pi/2$ (see an
illustration of these two geometries in Fig. \ref{fig:Geometry}).
Recalling the angular dependencies of the propagation coefficients,
$f \propto \cos \vartheta$ and $h \propto \sin^2 \vartheta$
[approximately, see equation (\ref{eq:h_rel})], this may result
in very different propagation effects, and hence in qualitatively
different polarization properties of the emitted radiation, as
we explain next.

In \S\,\ref{sec:Observational_Consequences_for_Uniform_Strong_Field}
we show that the frequency ranges where observational fingerprints
of plasma propagation effects are expected are $\nu \sim 1$ GHz for
the forward shock emission, and $\nu \sim 10^{13}$ Hz for
the reverse shock emission. Let us assume that the plasma is permeated by
a large scale magnetic field, and denote by $R_{f/h} = (f/h) \sin^2
\vartheta / \cos \vartheta $ the (geometry-independent) ratio
of the propagation coefficients in the relevant frequency ranges.
The plasma parameters we derive in
\S\,\ref{sec:Observational_Consequences_for_Uniform_Strong_Field}
imply that $R_{f/h} \sim 10 (\nu/1 \textrm{GHz})^{-1/3}$ in the
forward shock, and $R_{f/h} \sim 10^3 (\nu/10^{13} \textrm{Hz})$ in
the reverse shock~\footnote{The reason for the much larger value of
$R_{f/h}$ in the reverse shock is essentially the fact that reverse
shock radiation travels through a thick slab of \emph{cold} plasma,
whence the values of $f$ and $h$ follow from equation
(\ref{eq:f_and_h_cold}) rather than from eqs.
(\ref{eq:f_rel_delta_1}) and (\ref{eq:h_rel}), as they do in the
forward shock. The values we give here for $R_{f/h}$ are obtained
from plasma parameters of a fireball expanding into a uniform-density
ISM. Nevertheless, the corresponding values for a fireball expanding into
a wind are similar, and hence this scenario will not be treated
separately in this section.}.
Consider now a \emph{typical} jet with $\theta_j \gg \Gamma^{-1}$.
The high $R_{f/h}$ imply that whatever the direction of the magnetic
field be, in order for $h$ to dominate over $f$ (i.e. in order for
$\vartheta$ to be sufficiently close to $\pi/2$), $\mathbf{k}$ must
lie within a very narrow solid angle. Consequently, a configuration
where $f \gg h$ is much more probable than one in which $h \gg f$
[by a factor of $\sim R_{f/h} - 1 \sim 9\;\; (10^3)$ in the forward
(reverse) shock].
Next, consider a \emph{narrow} jet observed slightly off-axis, with
$\theta_j \sim \Gamma^{-1}$. Recall that in this regime, the vector
$\mathbf{k}$ is lying in the shock plane (see Fig.
\ref{fig:Geometry} \emph{right}). If the uniform magnetic field also
lies predominantly in the shock plane, we find again that $f \gg h$
is much more probable than $h \gg f$ [by factors of $\sim 7$ and
$\sim 10^{2.5}$ for the forward and reverse shocks, respectively].
If, however, $\mathbf{B}$ is normal to the shock plane, it is
necessarily normal also to $\mathbf{k}$, and therefore $h \gg f$.
The normal modes of propagating radiation are then linearly
polarized, and the resulting emission has markedly different
polarization properties.

In the next section we calculate propagation effects in both
geometric regimes. The case $f \gg h$ is studies in more detail,
since this is the characteristic case for a typical GRB jet.
The polarization pattern for this regime is shown in Fig.
\ref{fig:Combined_Forward_and_Reverse}. In Fig.
\ref{fig:Combined_Forward_and_Reverse_h_dom} we show the
polarization properties of the $h \gg f$ regime, which, as we
argued above, is probable only for a narrow jet observed off-axis,
with a uniform magnetic field normal to the shock plane.
The qualitative results of this discussion, complemented by some
quantitative consequences of the actual calculations, are
summarized in Table \ref{tbl:Dependence_on_Angle}.

\begin{deluxetable*}{ccccccc}[h!]
\tablecolumns{7}
\tabletypesize{\scriptsize}
\tablecaption{Effect of fireball, observing and field geometry on
the polarization\label{tbl:Dependence_on_Angle}}
\tablewidth{0pt}
\tablehead{
\colhead{} & \colhead{} &
\multicolumn{2}{c}{Uniform field} &
\colhead{} &
\multicolumn{2}{c}{Random field} \\
\cline{3-4} \cline{6-7} \\
\colhead{} & \colhead{} &
\colhead{$\perp$\phd\tablenotemark{a}} & \colhead{$||$\phd\tablenotemark{b}} &
\colhead{} &
\colhead{$\perp$} & \colhead{$||$} }
\startdata
$\theta_j \gg \Gamma^{-1}$ & {} & $f > h$ & $f>h$ & {} &
\multicolumn{2}{c}{$|f| > |h|$}  \\
{} & {} & Faraday depolarization & Faraday depolarization & {} &
\multicolumn{2}{c}{no propagation effects} \\
{} & \emph{FS\phd\tablenotemark{c}}~~: & $\Pi_C \gg \Pi_L$ at $\nu < 1$ GHz &
$\Pi_C \gg \Pi_L$ at $\nu < 1$ GHz & {} &
no polarization & $\Pi \lesssim 10$\% \\
{} & \emph{RS\phd\tablenotemark{d}}~~: & $\Pi_C \gg \Pi_L$ at
$\nu < \textrm{few} \times 10^{14}$ Hz &
$\Pi_C \gg \Pi_L$ at $\nu < \textrm{few} \times 10^{14}$ Hz & {} &
no polarization & $\Pi \lesssim 10$\% \\[6 mm]
$\theta_j \sim \Gamma^{-1}$ & {} & $h \gg f$ & $f > h$ & {} &
$|h| \gg |f|$ & $|f| > |h|$ \\
{} & {} & no Faraday depolarization & Faraday depolarization & {} &
\multicolumn{2}{c}{no propagation effects} \\
{} & \emph{FS}~~: & $\Pi_C \sim \Pi_L \sim 20\%-50\%$ at $\nu < 1$ GHz &
$\Pi_C \gg \Pi_L$ at $\nu < 1$ GHz & {} &
\multicolumn{2}{c}{$\Pi_C \ll \Pi_L \sim \Pi \lesssim \textrm{few} \times 10$\%} \\
{} & \emph{RS}~~: & $\Pi_C \ll \Pi_L \ll 100\%$ at $\nu < \textrm{few} \times 10^{13}$ Hz &
$\Pi_C \gg \Pi_L$ at $\nu < \textrm{few} \times 10^{14}$ Hz & {} &
\multicolumn{2}{c}{$\Pi_C \ll \Pi_L \sim \Pi \lesssim \textrm{few} \times 10$\%} \\[4 mm]
\enddata
\tablenotetext{a}{Field normal to plane of shock}
\tablenotetext{a}{Field parallel to plane of shock}
\tablenotetext{c}{Forward shock emission}
\tablenotetext{d}{Reverse shock emission}
\end{deluxetable*}



\section{Application to GRB afterglows}
\label{sec:Application_to_GRB_AG}

The application of the results derived in \S\,\ref{sec:Propagation_Effects}
to plasmas of GRB afterglows requires knowledge of the plasma number
density, strength and structure of the magnetic field, and the electron
distribution function produced by the shock. These parameters, in turn,
depend on the total energy $E$ carried by the relativistic plasma, the
number density $n$ of the ambient gas, the duration $T$ of the burst, the
Lorentz factor $\Gamma_i$ reached by the outflow prior to the deceleration
phase, the fractions $\epsilon_e$ and $\epsilon_B$ of the energy density
carried by relativistic electrons and magnetic field, respectively, and the
spectral index $p = -\textrm{d}\ln n_e / \textrm{d}\ln \gamma_e$ of the
assumed power-law distribution of the shock-accelerated electrons. When
considering a fireball expanding into a wind-stratified medium, $n$ is
replaced by $\dot{M} / 4\pi m_p R^2 v_{\rm w}$,
involving the ratio between the progenitor's mass loss rate, ${\dot M}$,
and the wind velocity, $v_{\rm w}$ (where $m_p$ is the mean atomic mass for
the gas). We shall therefore briefly address the observational constraints
on these parameters, on which our choices of parameter values is based.

\newcounter{f1}\stepcounter{f1}\stepcounter{f1}
We begin with the parameters characterizing the
explosion. First, we focus on the subclass of
long-duration ($T \gtrsim 2$ s) GRBs, since these are the only
bursts localized by the \emph{BeppoSAX} and \emph{HETE-\Roman{f1}}
missions with identified afterglows. We therefore express our
results in units of $T_1 = T / 10$~s.

Second, as we are interested in the emission prior to the jet break time,
the energy scale determining the dynamics is not the actual explosion
energy, but rather its isotropic equivalent, i.e.  the total energy derived
assuming isotropic emission.  Examination of the list of all bursts with
measured redshifts indicates $E_{\gamma,\rm iso} \simeq 3 \times 10^{53}$
ergs as the mean isotropic equivalent energy inferred from the $\gamma-$ray
fluence (Bloom, Frail \& Kulkarni 2003). Since this value must be smaller
than the total kinetic energy carried by the plasma, it is reasonable to
express $E_{\rm iso}$ in units of $10^{54}$ ergs.  Hereafter we drop the
subscript ``iso'' for brevity.

Considering now the ambient gas density, when the observed light curve can
be reconciled with expansion into a uniform density ISM, Bloom et
al. (2003) find $n$ to typically lie in the range $10^{0.5 \pm 1}
\textrm{cm}^{-3}$. We shall therefore adopt a parametrization with $n_0 = n
/ (1\, \textrm{cm}^{-3})$. On the other hand, it is now commonly believed
that the long duration GRBs are probably associated with core-collapse
supernovae. This association was originally motivated by evidence of
intense star formation in identified GRB host galaxies (Bloom, Kulkarni \&
Djorgovski 2002), and by additional evidence for optical supernovae
emission in several GRB afterglows (Bloom 2003); finally, the SN-GRB
relation was put on solid grounds with the recent detection of a supernova
emission spectrum (SN2003dh) associated with GRB030329 (Stanek et al. 2003;
Hjorth et al.  2003). Considering a likely phase of mass loss during the
late stages of the massive star's life, it is natural to examine a scenario
in which the fireball expands into a wind medium. Below we consider the
simplest wind model, where prior to the explosion the progenitor star
ejects mass at a constant rate $\dot{M} = 10^{-5}\, \dot{M}_{-5}
\textrm{M}_{\odot} \textrm{yr}^{-1}$ and at a constant speed $v_{\rm w} =
10^3 \, v_3 \textrm{km s}^{-1}$ (Chevalier \& Li 1999). These parameters
agree with observations of massive stars which are believed to be the
progenitors of SNe Ib/c associated with GRBs (Willis 1991). Note, that
since $n \sim 1~{\rm cm^{-3}}$ for a uniform-density circumburst gas is
inferred from observations carried typically $\sim 10$ hours after the
burst, this is consistent also with a wind environment, which should have
comparable density at the radius reached by the fireball on a $\sim 10$
hours time scale (Livio \& Waxman 2000). Finally, the Lorentz factor
reached by the outflow prior to its deceleration by the circumburst gas is
typically a few hundreds. We choose to parameterize our results in
units of $\Gamma_{i,2.5} = \Gamma_i / 10^{2.5}$.

We now turn to consider the parameters $p$, $\epsilon_e$ and
$\epsilon_B$. There is strong indication that the $p$ and $\epsilon_e$ are
uniform among different bursts, with $\epsilon_e$ close to an equipartition
value (Frail et al. 2001; Freedman \& Waxman 2001; Berger, Kulkarni \&
Frail 2003), and $p = 2.2 \pm 0.1$ (Waxman 1997a; Galama et al. 1998; Stanek
et al. 1999; Frail, Waxman \& Kulkarni 2000; Freedman \& Waxman 2001). We
therefore adopt $\epsilon_e = 0.1 \epsilon_{e,-1}$ and $p = 2.2$ in the
numerical examples we consider below. The value of $\epsilon_B$ is less
constrained by observations, and its estimated value ranges from
$\epsilon_B \sim 10^{-1}$ (e.g., Waxman 1997a; Wijers \& Galama 1999) to
$\epsilon_B \sim 10^{-6}$ (e.g., Wijers \& Galama 1999; Chevalier \& Li
1999; Galama et al. 1999; Waxman \& Loeb 1999). Since in cases where
$\epsilon_B$ can be reliably constrained by multi-waveband spectra, values
close to equipartition are inferred (Frail et al. 2000), we first explore
consequences of propagation effects on emission from plasmas with
$\epsilon_B = 0.1 \epsilon_{B,-1}$. This is also the parametrization used
in \S\,\ref{sec:Plasma_Parameters}. We then study the implications of a
weak magnetic field in \S\,\ref{sec:Magnetic_Field_Strength_and_Structure},
adopting $\epsilon_B = 10^{-4}$ as an illustrative example.

Next we explore how the propagation effects described in the previous
section manifest themselves in the context of a GRB afterglow. In
\S\,\ref{sec:Plasma_Parameters} we derive the relevant plasma conditions at
the onset of fireball deceleration, considering both expansion into a
uniform density ISM (\S\,\ref{sec:Plasma_Parameters_ISM}) and into a wind
medium (\S\,\ref{sec:Plasma_Parameters_Wind}). In
\S\,\ref{sec:Observational_Consequences_for_Uniform_Strong_Field} we
examine the observable consequences of our derivations, investigating first
the case of a uniform, close to equipartition magnetic field.
Since a typical GRB jet is characterized by $\theta_j \gg \Gamma^{-1}$,
implying typically $f \gg h$ in the frequency ranges of interest (see
\S\,\ref{sec:Field_Direction_and_Viewing_Geometry}), this section gives
detailed analysis of this regime. \S\,\ref{seq:Narrow_Jet}
studies the complementary case, where $h \gg f$.
The effect of a weak field, as well as the effect of an entangled field
geometry, are treated in \S\,\ref{sec:Magnetic_Field_Strength_and_Structure}.
Throughout the rest of this paper, we use primed quantities to denote values
measured in the frame that is comoving with the plasma.

\subsection{Plasma parameters during the early afterglow}
\label{sec:Plasma_Parameters}

\subsubsection{Expansion into a uniform-density ISM}
\label{sec:Plasma_Parameters_ISM}

The relativistic blast wave driven into the ambient gas by the energetic
explosion approaches a self-similar behavior (Blandford \& McKee 1976) once
a reverse shock crosses the ejecta and heats it (M\'esz\'aros \& Rees
1997). During this transition episode both the shocked ambient gas and the
heated ejecta propagate with a Lorentz factor which is close to that given
by the Blandford-McKee self-similar solution (see, e.g. Waxman \& Draine
1999)~:
\begin{equation}\label{eq:Gamma_FR_ISM}
\Gamma^{(R,F)} \simeq \left( \frac{17 E}{1024 \pi n m_p \, c^5 \,
T^3} \right)^{1/8} \simeq 328\; E_{54}^{1/8} n_{0}^{-1/8}
T_1^{-3/8},
\end{equation}
where $E_{54} = E / 10^{54}\, \textrm{erg}$ is the isotropic-equivalent of
the fireball energy, $T_1 = T / 10\, \textrm{s}$ is the burst duration, and
$n_0 = n / 1\, \textrm{cm}^{-3}$ is the ambient number density.
The superscripts $R,F$ denote the reverse and forward shocked plasmas,
respectively.
The corresponding electron number densities are accordingly given by
the shock jump conditions~:
\begin{eqnarray}\label{eq:n_e_FR_ISM}
n_e^{\prime\,(F)} &=& 4 \Gamma n \simeq 1300\; E_{54}^{1/8}
n_{0}^{7/8} T_1^{-3/8}\;
\textrm{cm}^{-3} \nonumber\\
n_e^{\prime\,(R)} &\simeq& (\Gamma^2 / \Gamma_i) n_e^{\prime\,(F)}
\nonumber\\
&\simeq& 3.99 \times 10^5\; E_{54}^{3/8} n_{0}^{5/8} T_1^{-9/8}
\Gamma_{i,2.5}^{-1} \;\textrm{cm}^{-3}\;.
\end{eqnarray}

The comoving widths of shocked forward and reverse plasma shells
can be shown to be comparable, and are estimated as~:
\begin{equation}\label{eq:W_FR_ISM}
W^{\prime\,(F,R)} \simeq R / 4\Gamma \simeq 9.8 \times 10^{13}\;
E_{54}^{1/8} n_{0}^{-1/8} T_1^{5/8}\; \textrm{cm}\;,
\end{equation}
where $R \simeq 1.3 \times 10^{17}$ cm is the radius where the reverse
shock is propagating through the ejecta. From the shock jump conditions we
infer the strength of the magnetic field~:
\begin{eqnarray}\label{eq:B_FR_ISM}
B^{\prime\,(F,R)} &=& \left[ 32 \pi \epsilon_B n m_p c^2
\right]^{1/2} \Gamma \nonumber\\
&\simeq&  40\; E_{54}^{1/8} n_{0}^{3/8}
T_1^{-3/8} \epsilon_{B,-1}^{1/2}\; \textrm{G}\;.
\end{eqnarray}
The distribution function of relativistic electrons injected by
the shocks is assumed to increase as $\gamma_e^2$ up to a
characteristic Lorentz factor $\gamma_m$ (Gruzinov \& Waxman
1999), and decrease as a power law $\gamma_e^{-p}$ (with p = 2.2)
at higher values of the Lorentz factor. Then the jump conditions
imply~:
\begin{eqnarray}\label{eq:gamma_FR_ISM}
\gamma_m^{\prime\,(F)} &=& \frac{4(p-2)}{3(p-1)} \epsilon_e
\frac{m_p}{m_e} \Gamma \nonumber\\
&\simeq& 1.32 \times 10^4\; E_{54}^{1/8} n_{0}^{-1/8} T_1^{-3/8}
\epsilon_{e,-1}\; \nonumber\\
\gamma_m^{\prime\,(R)} &=& \frac{4(p-2)}{3(p-1)} \epsilon_e
\frac{m_p}{m_e}  \frac{\Gamma_i}{\Gamma} \nonumber\\
&\simeq& 43\; E_{54}^{-1/8}
n_{0}^{1/8} T_1^{3/8} \Gamma_{i,2.5} \epsilon_{e,-1}\;.
\end{eqnarray}

The observed characteristic synchrotron frequencies corresponding
to the Lorentz factor of the thermal electrons in these two shock
scenarios are~:
\begin{eqnarray}\label{eq:Important_frequencies_FR_ISM}
\nu_{m}^{(F)} &\simeq& 9.61 \times 10^{18}\; E_{54}^{1/2}
T_1^{-3/2} \epsilon_{e,-1}^2 \epsilon_{B,-1}^{1/2}\;
\textrm{Hz}\nonumber\\
\nu_{m}^{(R)} &\simeq& 1.04 \times 10^{14}\; n_0^{1/2}
\Gamma_{i,2.5}^2 \epsilon_{e,-1}^2 \epsilon_{B,-1}^{1/2}\;
\textrm{Hz};
\end{eqnarray}
In the parlance of the high-energy and IR communities,
respectively, these read an energy $E_{ph,m}^{(F)} =
39.7$ KeV for the forward shock, and a wavelength
$\lambda_m^{(R)} = 2.9 \mu$m for the reverse shock.

\subsubsection{Expansion into a wind}
\label{sec:Plasma_Parameters_Wind}

We now examine a fireball expanding into a wind. The simple wind
model considered here results in a non-homogeneous ambient gas
with a profile $n \propto R^{-2}$. The reverse shock crosses the
ejecta at a typical radius of $2.2 \times 10^{16}$ cm, at which
time the plasmas shocked by the forward and reverse shocks
propagate together with a Lorentz factor
\begin{equation}\label{eq:Gamma_FR_Wind}
\Gamma^{(F,R)} \simeq 135\; E_{54}^{1/4} (\dot{M}_{-5}/v_3)^{-1/4}
T_1^{-1/4}\;.
\end{equation}
Following the reasoning employed above, we obtain the other
relevant plasma parameters for the wind scenario~:
\begin{eqnarray}\label{eq:Plasma_Parameters_FR_wind}
W^{\prime\,(F,R)} &\simeq& 4.04 \times 10^{13}\; E_{54}^{1/4}
(\dot{M}_{-5}/v_3)^{-1/4} T_1^{3/4}\; \textrm{cm}\nonumber\\
n_e^{\prime\,(F)} &\simeq& 3.43 \times 10^5 E_{54}^{-3/4}
(\dot{M}_{-5}/v_3)^{7/4} T_1^{-5/4}\;\textrm{cm}^{-3} \nonumber\\
n_e^{\prime\,(R)} &\simeq& 1.78 \times 10^7 E_{54}^{-1/4}
(\dot{M}_{-5}/v_3)^{5/4} T_1^{-7/4} \Gamma_{i,2.5}^{-1}\;
\textrm{cm}^{-3} \nonumber\\
B^{\prime\,(F,R)} &\simeq& 418\; E_{54}^{-1/4}
(\dot{M}_{-5}/v_3)^{3/4} T_1^{-3/4} \epsilon_{B,-1}^{1/2}\;
\textrm{G} \nonumber\\
\gamma_m^{\prime\,(F)} &\simeq& 5440\; E_{54}^{1/4}
(\dot{M}_{-5}/v_3)^{-1/4} T_1^{-1/4} \epsilon_{e,-1}\;
\nonumber\\
\gamma_m^{\prime\,(R)} &\simeq&105\; E_{54}^{-1/4}
(\dot{M}_{-5}/v_3)^{1/4} T_1^{1/4} \Gamma_{i,2.5}
\epsilon_{e,-1}\;.
\end{eqnarray}

The characteristic synchrotron frequencies in the wind scenario
are~:
\begin{eqnarray}\label{eq:Important_frequencies_FR_Wind}
\nu_{m}^{(F)} &\simeq& 7.0 \times 10^{18}\; E_{54}^{1/2}
T_1^{-3/2} \epsilon_{e,-1}^2 \epsilon_{B,-1}^{1/2}\;
\textrm{Hz}\;,\nonumber\\
\nu_{m}^{(R)} &\simeq& 2.60 \times 10^{15}\; E_{53}^{-1/2}
(\dot{M}_{-5}/v_3) T_1^{-1/2} \Gamma_{i,2.5}^2 \nonumber\\
&\qquad\qquad& \times\epsilon_{e,-1}^2\epsilon_{B,-1}^{1/2}\; \textrm{Hz}.
\end{eqnarray}

\subsection{Observational consequences for a uniform,
close to equipartition magnetic field ---
the ``typical jet'' regime }
\label{sec:Observational_Consequences_for_Uniform_Strong_Field}

The formalism developed in \S\,\ref{sec:Brief_Formalism} was
applied to the plasma conditions derived above. We assumed the
magnetic field and electron density are uniform across the slab.
As an electron drifts away from the shock front with a comoving
velocity $c/3$, it loses energy due to synchrotron emission.
Consider, for simplicity, an isotropic, monoenergetic  electron
population, with a Lorentz factor $\gamma_{i}^\prime$ endowed
by the shock. Assuming this population remains isotropic (in
momentum space of the comoving plasma) as it loses energy to
radiation, the mean Lorentz factor decreases as
\begin{equation}\label{eq:Cooling_of_gamma}
\gamma^\prime(z^\prime) = \frac{\gamma_{i}^\prime} {1 + 3 \sigma_T
{B^\prime}^2 \gamma_{i}^\prime z^\prime / 6 \pi m_e c^2}\;,
\end{equation}
where $z^\prime$ is the comoving separation from the shock front.
Assuming now that the shock injects a spectrum of electron energies,
equation (\ref{eq:Cooling_of_gamma}) implies that the shape of the
\emph{energy} distribution function is modified as the separation
from the shock front increases. As a result, only a thin slab of
width $3 \times 10^{11}$ cm behind the forward shock is dominated
by a relativistic electron population, while the reverse shock front
is separated from the observer by a cold slab $2 \times 10^{13}$
cm wide (for a fireball expanding into ISM; The corresponding widths
for the wind scenario are $5 \times 10^{9}$ cm and $4 \times 10^{13}$
cm, respectively).
As this strongly affects the values of emission,
absorption and propagation coefficients (see Appendix
\ref{sec:appendix_A}), we integrated the radiative transfer equation
(\ref{eq:Transfer_Equation_for_Stokes_Parameters}) by dividing the
emitting slab into a large number of sub-slabs, such that within
each sub-slab the electron energy spectrum did not change
significantly. The coefficients of emission, absorption and
propagation could then be assumed constant within each sub-slab,
allowing an analytic solution of the transfer equation.
This division was refined until convergence was obtained.

Equations (\ref{eq:f_and_h_cold}), (\ref{eq:f_rel_delta_1}) and
(\ref{eq:h_rel}) imply a strong dependence of the propagation
coefficients on $\vartheta$, the comoving angle between the
directions of the wave-vector and the magnetic field.
Yet, as argued in \S\,\ref{sec:Field_Direction_and_Viewing_Geometry},
the observing geometry for a typical GRB jet are likely to give
rise to a situation where $f \gg h$. In such circumstances
$\vartheta \sim \pi/2$ is most unlikely, and the angular dependence
of $f$ and $h$ may be ignored, for simplicity. This simplification
was adopted in the calculation descried below, and gives a good
order of magnitude estimate of the emitted polarization of a
typical jet with probable viewing geometry.
The case $h \gg f$, occurring when the jet is narrow ($\theta_j \sim
\Gamma^{-1}$) and the magnetic field is predominantly normal to the
shock plane (see \S\,\ref{sec:Field_Direction_and_Viewing_Geometry})
is treated separately in \S\,\ref{seq:Narrow_Jet}.

We verified that all elements of the tensor $\kappa_{ij}$ at all relevant
frequencies are much smaller than unity, thus self-consistently justifying
the analysis of \S\,\ref{sec:Brief_Formalism} which assumed that the
electric field is perpendicular to the wave vector.

\begin{figure}
\epsscale{0.9}
\plotone{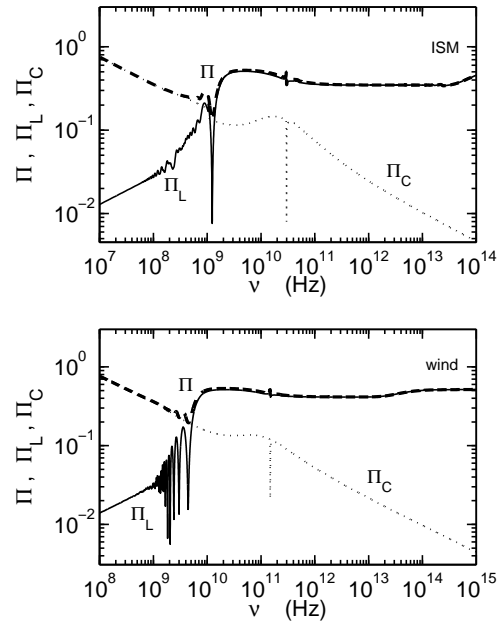}
\caption{Propagation effects on synchrotron radiation propagating
in a magnetoactive plasma. The plasma conditions are
characteristic of \emph{forward} shocks for a fireball expanding
into a uniform-density ISM (\emph{above}) and into wind
(\emph{below}), during the onset of fireball deceleration. The
magnetic field is assumed uniform and close to equipartition
($\epsilon_B = 10^{-1}$). Both figures show the degrees of linear
polarization $\Pi_L$ (\emph{solid}), circular polarization
$\Pi_C$ (\emph{dotted}), and the total degree of polarization $\Pi$
(\emph{dashed}). \label{fig:Forward_Shocks}}
\end{figure}

\begin{figure}
\epsscale{0.9}
\plotone{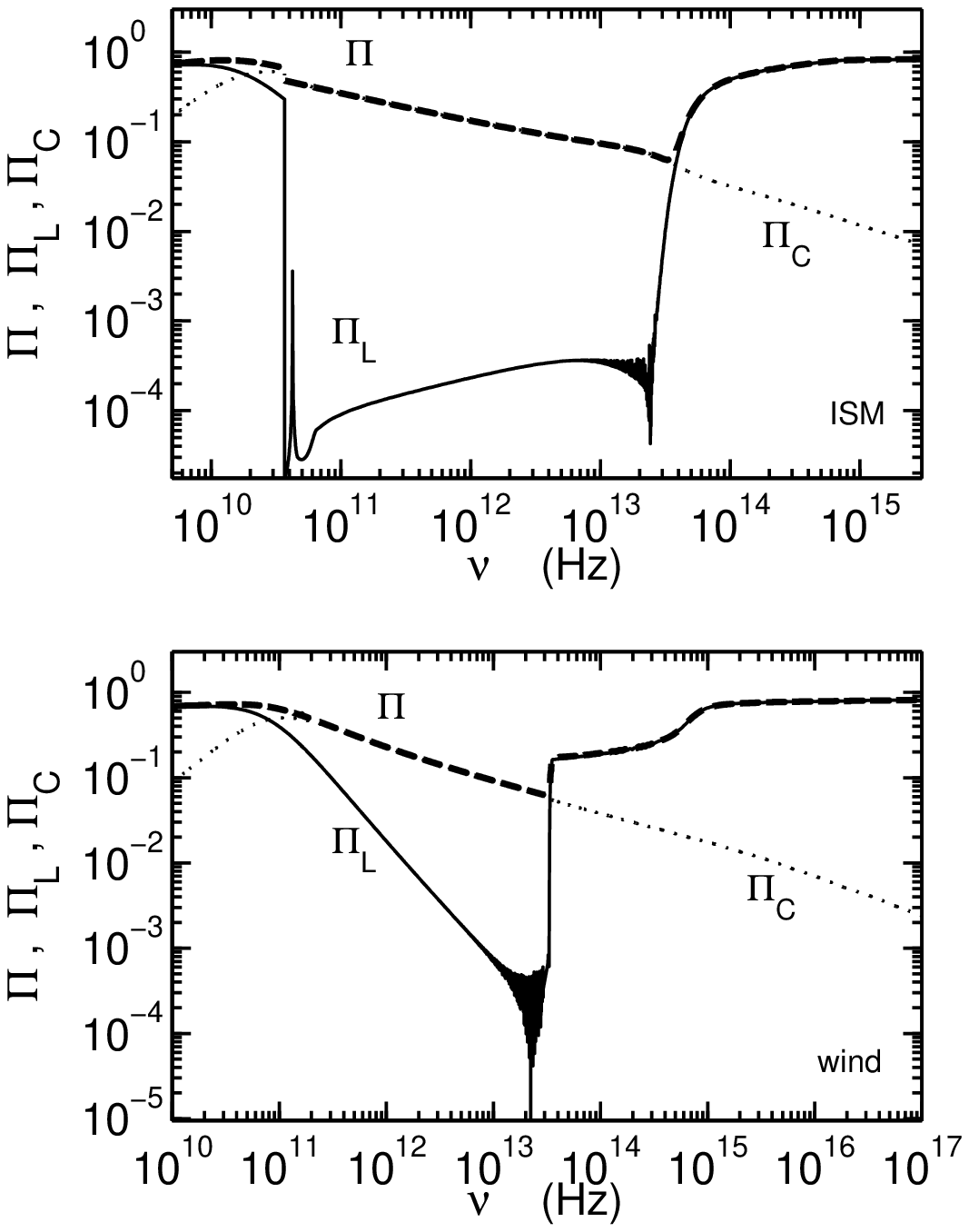}
\caption{Propagation effects on synchrotron radiation propagating
in a magnetoactive plasma. The plasma conditions are
characteristic of \emph{reverse} shocks for a fireball expanding
into a uniform-density ISM (\emph{above}) and into wind
(\emph{below}), during the onset of fireball deceleration. The
magnetic field is assumed uniform and close to equipartition
($\epsilon_B = 10^{-1}$). Both figures show the degrees of linear
polarization $\Pi_L$ (\emph{solid}), circular polarization
$\Pi_C$ (\emph{dotted}), and the total degree of polarization $\Pi$
(\emph{dashed}). \label{fig:Reverse_Shocks}}
\end{figure}

The resulting emitted linear, circular and total degrees of polarization
at the onset of fireball deceleration are shown in Figs.
\ref{fig:Forward_Shocks} (forward shock) \& \ref{fig:Reverse_Shocks}
(reverse shock).
We show results for both ISM and wind
scenarios. All four graphs exhibit broadly the same behavior. Due to the
Faraday effect, the linear polarization of emitted radiation is strongly
suppressed at low frequencies, and circular polarization dominates. At a
frequency of $\sim 1$ GHz (30 GHz) in the forward shock into ISM (wind),
and $\sim 3 \times 10^{13}$ Hz in the two reverse shocks, linear
polarization takes over. At the transition frequency the total degree of
polarization has a global minimum (10-20\%). Only at frequencies much
higher than this transition frequency the canonical degree of polarization
(dominated by its linear component) is reached.

The emission from the reverse shock overtakes the forward shock
emission at frequencies above $\nu \sim 2 \times 10^{10}$ Hz. We
find that for the reverse shock, the transition from circular to
linear polarization is accompanied by a strong oscillation of the
polarization position angle as function of frequency. Due to
synchrotron losses, most of the plasma heated by the reverse shock
cools quickly on a dynamic time scale, and consequently Faraday
rotation by the cold plasma strongly affects radiation emitted
from regions closer to the reverse shock front. The resulting
oscillation is a hallmark of reverse shock emission up to UV
frequencies ($10^{15}-10^{16}$ Hz), and is characterized by
$\Delta\nu/\nu \simeq 10^{-1} \nu_{15}^2$ [from equation
(\ref{eq:Delta_nu_over_nu})], where $\nu_{15}$ is the observed
frequency in units of $10^{15}$ Hz;
Observation of these oscillations may therefore be limited by
instrument resolution in the optical and particularly the IR
bands. Note, additionally, that during the transition to
self-similar dynamics, emission is observed from an area on the
sky with a radius comparable to $R/\Gamma$ (Waxman 1997b). Hence
radiation at observed frequency $\nu$ spans a range of
\emph{comoving} frequencies $\Delta\nu^\prime$. Since radiation at
different comoving frequencies is subject to different phase
shifts between normal modes, spectro-polarimetric features
characterized by a smaller comoving $\Delta\nu^\prime$ may be
washed out altogether. Although an exact, quantitative statement
regarding the implications to the observability of the Faraday
oscillations is beyond the scope of this paper, we expect that
the depolarization of the linear component, caused by either low
spectral resolution or smearing of the signal by the source plasma,
is expected to yield a signal dominated by a circular polarization
$\Pi_C \sim 0.1 (\nu / 10^{13} \textrm{Hz})^{-1/3}$, surviving up
to UV frequencies.

\begin{figure}
\epsscale{0.9}
\plotone{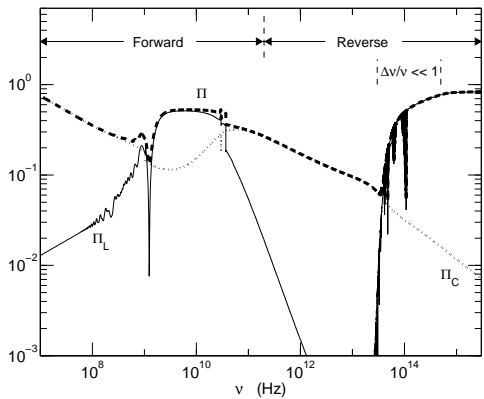}
\caption{Polarization pattern of combined (superposed) reverse
and forward shock emission, for a fireball expanding into a
uniform-density ISM. Shown are the degrees of linear polarization
$\Pi_L$ (\emph{solid}), circular polarization $\Pi_C$ (\emph{dotted}),
and the total degree of polarization $\Pi$ (\emph{dashed}).
At frequencies $\nu < 2 \times 10^{11}$ Hz the polarization
is determined by forward shock emission, whereas above that
frequency it is dominated by the emission from the reverse shock.
In the frequency range $3 \times 10^{13}\, \textrm{Hz} < \nu
\lesssim \textrm{few} \times 10^{14} \, \textrm{Hz}$ marked by
$\Delta\nu \ll \nu$ in the figure, rapid oscillations of the
polarization position angle may not be resolved, thus suppressing
linear polarization, rendering circular polarization dominant.
The calculations leading to this result correspond to a typical
GRB jet with an opening angle $\theta_j \gg \Gamma^{-1}$, implying
$f \gg h$ (see \S\,\ref{sec:Field_Direction_and_Viewing_Geometry}).
\label{fig:Combined_Forward_and_Reverse}}
\end{figure}

A superposition of the polarization patterns of the forward and
reverse shocks is shown in Fig.
\ref{fig:Combined_Forward_and_Reverse}, for the ISM scenario.
The polarization is dominated by the reverse shock emission at
frequencies above $\sim 2 \times 10^{11}$ Hz. The analogous
superposition of forward and reverse shock polarization
for the wind scenario is qualitatively similar.

Remarkably, the main characteristics of the polarization pattern, including
the frequency of transition from circular to linear polarization, are
largely independent of the circumburst density, and are dictated
predominantly by the nature of the shock (i.e. whether the shock is forward
or reverse). This is particularly intriguing in view of the three orders of
magnitude factor separating the densities of ambient gas in the ISM and
wind scenarios, at the onset of fireball deceleration. Considering the two
forward shocks first, it is straightforward to show that in both cases
$\nu_a^\prime \ll \nu_m^\prime$, where $\nu_a^\prime$ is the (comoving)
synchrotron self-absorption frequency. At smaller frequencies, $\nu^\prime
< \nu_a^\prime$, the effective width $W^\prime_{eff}(\nu^\prime)$ from
which photons are emitted scales as $W^\prime_{eff}(\nu^\prime) \propto
[\kappa^\prime(\nu^\prime)]^{-1} \propto {\nu^\prime}^{5/3}$. By virtue of
equations (\ref{eq:Pi_L}) and (\ref{eq:f_rel_delta_1}), then, we have at
low frequencies $\Pi_L(\nu^\prime) \propto W^\prime_{eff} {\nu^\prime}^2
\propto {\nu^\prime}^{1/3}$. On the other hand, the degree of circular
polarization at frequencies $\nu^\prime \gg \nu_B^\prime$ scales as
${\nu^\prime}^{-1/3}$.  Denoting the frequency of the transition from
circular to linear polarization by $\nu_t$ we therefore find $\nu_t^\prime
\propto (\nu_B^\prime \nu_a^\prime)^{1/2}$. Transforming to the observer's
frame, we find for the forward shock in the ISM, $\nu_t \propto n^{27/80}
\sim n^{1/3}$. Substituting here the ambient density for the wind scenario
at the transition-phase radius, which is larger by $\sim$ 3 orders of
magnitudes, and correcting for the much narrower slab which effectively
emits photons in this scenario (due to the more efficient synchrotron
cooling), we obtain $\nu_t$ in the wind scenario larger by a factor of
$\sim 3$, as can be readily verified in Figure
\ref{fig:Forward_Shocks}. Considering now the two reverse shocks, the
frequency at which linear polarization takes over circular polarization is
simply the synchrotron self-absorption frequency, since at $\nu > \nu_a$
the region closest to the shock front becomes observable. As this region is
populated by electrons which are still highly relativistic, it contributes
most of the photons above $\nu_a$, while the thick slab of cold plasma
separating this emitting region from the observer is characterized by
a much smaller emissivity. Radiation emerging from the emitting region
close to the reverse shock front is subject to a large Faraday
rotation due to the presence of the cold plasma; however, since the
emissivity of the cold plasma is small, this has only a minor effect
on the degree of linear polarization of that radiation.
The similar
$\nu_t$ in the two reverse shock scenarios is thus a consequence of a
similar self-absorption frequency in these two cases.

We note that our calculations indicate the plasma shocked by the forward
shock does not affect the polarization properties of the emission from the
reverse shock, since it is optically thin in the frequency range where the
reverse shock emission exhibits interesting polarization features.
Moreover, only the non-relativistic component of the plasma behind the
forward shock introduces a Faraday rotation with a large enough amplitude
above a frequency of $10^{13}$ Hz. Yet this plasma contributes only
negligibly to the emission and absorption of radiation, and hence does not
change the linear, circular or total degrees of polarization, but merely
modifies weakly the pattern of position angle oscillations.

\subsection{The ``narrow jet'' regime --- observational consequences}
\label{seq:Narrow_Jet}

\begin{figure}
\epsscale{1} \plotone{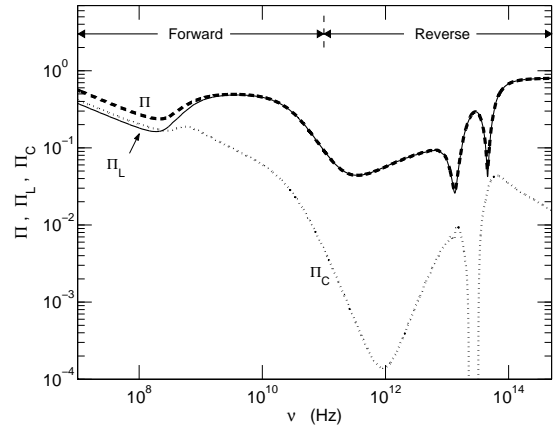}
\caption{Same as Fig. \ref{fig:Combined_Forward_and_Reverse}, but
with $\vartheta$ set to $\pi/2$, implying effectively $h \gg f$.
This situation is a-typical, and is probable for a narrow jet of
an opening angle $\theta_j \sim \Gamma^{-1}$, with the jet observed
off-axis, and the uniform magnetic field directed normal to the
shock plane.
Plasma parameters correspond to a fireball expanding into ISM.
However, we find that the picture is qualitatively similar also
in the wind scenario.
\label{fig:Combined_Forward_and_Reverse_h_dom}}
\end{figure}

A narrow jet ($\theta_j \sim \Gamma^{-1}$) observed slightly off-axis,
with a uniform magnetic field normal to the plane of the shock(s) is
the only geometric configuration where the propagation coefficient
$h$ is very likely to dominate over $f$ (see
\S\,\ref{sec:Field_Direction_and_Viewing_Geometry}).
We repeated the calculations of plasma propagation effects on the
polarization properties of early afterglow emission, assuming
uniform, close to equipartition fields ($\epsilon_B = 0.1$) permeate
the reverse and forward shocked plasmas, but this time neglecting
$f$ in comparison to $h$ [i.e., setting $\vartheta = \pi/2$ in
equations (\ref{eq:f_and_h_cold}) - (\ref{eq:h_rel})].
The resulting linear, circular and total degrees of polarization
of the combined forward-reverse shock emission are shown in Fig.
\ref{fig:Combined_Forward_and_Reverse_h_dom}.
The emerging picture is markedly different than the one described
above (\S\,\ref{sec:Application_to_GRB_AG}, Fig.
\ref{fig:Combined_Forward_and_Reverse}).
Linear polarization is not suppressed by Faraday rotation, and
dominates over circular polarization in the entire frequency
range of interest. (Notice, however, the comparable degree of
circular polarization in the emission from the forward shock at
low frequencies, $\nu \lesssim \textrm{few} \times 100$ MHz, due
entirely to the intrinsic circular polarization of synchrotron
at low frequencies). Additionally, we notice the absence of
Faraday oscillations of the polarization position angle $\chi$
at high (IR to UV) frequencies. Not being ironed out by the
oscillations, linear polarization is therefore expected to prevail
in this case in the IR - UV range as well, in contrast to the
situation in the case of a typical jet.
Interestingly, linear polarization \emph{is} suppressed in the
microwave and IR band (contributed solely by reverse shock emission)
not by the action of plasma effects, but rather by the suppression
of the Stokes parameter $Q$ relative to $I$ at large optical
depths~\footnote{See, e.g., Pacholczyk (1977) for an analogous
suppression of $\Pi_L$ at $\nu \gg \nu_m$}.

\subsection{Implications of magnetic field strength and structure}
\label{sec:Magnetic_Field_Strength_and_Structure}

As mentioned above, the fraction of the post-shock energy density carried
by the magnetic field, $\epsilon_B$, is one of the model parameters least
constrained by observations. Since the propagation effects addressed in
this paper are particularly sensitive to the strength of the magnetic
field, it is natural to explore how the polarization of emitted radiation
depends on the amplitude of $B$.  We therefore consider $\epsilon_B =
10^{-4}$ as a fiducial value for the weak field regime, keeping all other
fireball parameters unchanged.

We find that the general features exhibited by the linear,
circular and total degrees of polarization are broadly similar to
those described above for the uniform equipartition field scenario
described above. In particular, we find that the linear
polarization is suppressed at low frequencies. There are some
marked differences, though~: the transition frequency from
circular to linear polarization is decreased by approximately an
order of magnitude compared to the strong field scenarios
described above. Also, the oscillation pattern of the position
angle in the reverse shock scenario does not survive the decrease
in $B$, and is absent in the weak field regime.

In the case where the structure of the magnetic field is entangled
over length scales $\ell$ much shorter than the typical width of
the emitting slabs $W$, the fluctuating phase accumulated between
the two normal modes averages to zero, and its r.m.s. is suppressed
by a factor $(\ell/W)^{1/2} \ll 1$, resulting in the absence of the
propagation effects discussed above. It is helpful in this context
to separate the discussion of the two fireball and observation
geometric regimes. Addressing first the \emph{typical jet} regime
($\theta_j \gg \Gamma^{-1}$) in the random field scenario ($\ell \ll
W$), we conclude that linear polarization is not subject to suppression
due to the Faraday rotation. Nevertheless, this does not imply a high
degree of linear polarization, because the overall orientation of
synchrotron radiation emitted from the many different patches averages
to zero. If, however, the scale of the coherent patch is limited only
by causality and is thus comparable to the width of the emitting slab,
viz. $\ell \sim W$ (Gruzinov \& Waxman 1999), Faraday rotation is not
completely suppressed, and some suppression of $\Pi_L$ with respect
to $\Pi_C$ is expected. The frequencies of transition from circular to
linear polarization are not expected to change substantially, although
the degrees of circular and linear polarization will be lower by a
factor of a few (with a maximal $\Pi$ at the level of $\sim 10$\%),
due to the different directions that the magnetic field may assume
in the $\sim 50$ coherent, causal patches apparent to the observer
(Gruzinov \& Waxman 1999).
Turning our attention to the \emph{narrow jet} regime now, a random,
$\ell \ll W$ field will suppress all propagation effects. In this
case a high degree of linear polarization --- $\Pi_L \sim \textrm{few}
\times 10$\% --- may arise, if the random magnetic field has a dominant
(fluctuating) component either normal to the plane of the shock or
parallel to it (Waxman 2003b; Nakar et al. 2003). The high polarization
stems from the orthogonality of the direction $\mathbf{k}$ of observed
radiation to the direction of the magnetic field in the frame comoving
with the plasma (see Fig. \ref{fig:Geometry} \emph{right} for
illustration). Independent patches limited by causality, like in
Gruzinov \& Waxman (1999), will not suppress propagation effects
altogether, and the polarization pattern of the emitted radiation
is expected to follow qualitatively either Fig.
\ref{fig:Combined_Forward_and_Reverse} or Fig.
\ref{fig:Combined_Forward_and_Reverse_h_dom}, depending on the dominant
component of the fluctuating magnetic field, with maximum levels of
$\Pi_L$ and $\Pi_C$ decreased by a factor of few. These predictions
are summarized in Table \ref{tbl:Dependence_on_Angle}.

To conclude, then, the polarization of the early afterglow emission
provides a robust measure of the field's complexity~: A dominant, high
degree of circular polarization at $\nu \lesssim 1$ GHz and again at
$\textrm{few} \times 10^{10}\,\, \textrm{Hz} \lesssim \nu \lesssim
\textrm{few} \times 10^{14}\,\, \textrm{Hz}$
is a clear signature of a large scale, coherent field present in the plasma.
On the other hand, a high degree of linear polarization may result with
high probability in the narrow jet regime both from a large scale field
normal to the shock plane, or from an entangled field, and is therefore
less informative as far as field structure is concerned.


\section{Conclusion}
\label{sec:Conclusion}

Ever since it was realized that the dominant mechanism responsible for the
observed radiation from GRB afterglows is synchrotron emission, the spectra
and polarization at various stages of fireball expansion were calculated by
integration of the relevant emissivities and absorption coefficients over
varying conditions in the fireball plasma. Indeed, lightcurves and spectra
are successfully reproduced by such procedures. However, the propagation
effects of a magnetized, relativistic plasma on the polarized synchrotron
radiation may have significant observable outcomes, and were not treated so
far. In particular, the birefringent nature of the magnetized plasma,
leading to a phase shift between the normal propagation modes, which is
accumulated on the coherence length scale of the magnetic field, and
manifested as the Faraday effect, may bear significant consequences for
the polarization of GRB afterglows; A full treatment of this problem,
taking into consideration the central part played by synchrotron
cooling (see below), is presented here for the first time.

In this work we generalized results by Sazonov (1969), and calculated the
terms of the dielectric susceptibility tensor responsible for the Faraday
effect for a plasma with an isotropic, relativistic distribution of
electrons. We also derived new simple, analytic expressions for the
propagation coefficients in various frequency ranges (see
Eqs. [\ref{eq:f_rel_delta_1}], [\ref{eq:h_rel}], and Appendix
\ref{sec:Propagation_Coefficients_Appendix}). Similar expressions were
derived independently by Matsumiya \& Ioka (2003).

We have applied these results to plasma conditions typical of the early
afterglow, where plasma effects on the polarization of propagating
synchrotron radiation are significant. We considered both the forward and
reverse shocks, and showed that if the plasma is permeated by a
large-scale, close to equipartition magnetic field, the polarization
pattern of emitted radiation differs decisively from the naive estimates
usually quoted in the literature.
We showed that for a typical GRB jet, in which $\theta_j \gg \Gamma^{-1}$
during the early afterglow, a large-scale field results in a strong
suppression of linear polarization at low frequencies, i.e. $\nu \la 3$
GHz ($\nu<3\times 10^{13}$ Hz) in a forward (reverse) shock due to
Faraday rotation, and a prevalence of circular polarization.
Above these transition frequencies linear polarization takes over, and
gradually approaches the canonical value. At the frequencies of transition,
the total degree of polarization obtains a global minimum of 10-20\%.
The emission from the reverse shock above the characteristic
synchrotron frequency exhibits strong, fast oscillations of the polarization
position angle as a function of frequency. The oscillations mark the
emission up to UV frequencies ($10^{15} - 10^{16}$ Hz); since they
may be spectrally resolved only above $\sim \textrm{few} \times 10^{14}$
Hz, this implies a suppression of linear polarization (and a dominating
circular polarization as a result) up to the UV band.
These prominent fingerprints of the presence of a large-scale magnetic
field in the plasma are rather independent of the environment into
which the fireball expands, whether it has a uniform density (ISM)
or a wind density profile (see
\S\,\ref{sec:Observational_Consequences_for_Uniform_Strong_Field}).
The suppression of linear polarization at low frequencies was shown to
characterize also the case of a uniform field which is much below the
equipartition value. However, in the weak field regime the frequency of
transition from circular to linear polarization decreases by
approximately an order of magnitude, and the oscillation pattern of the
position angle is absent (whence also the dominance of $\Pi_C$ up to the
UV band). These propagation effects are expected to vanish if the field
is entangled over length scales much smaller than the extent of the
emitting plasma.

A large-scale magnetic field is likely to give rise to different
dispersive effects on the polarization properties if the jet is narrow
($\theta_j \sim \Gamma^{-1}$) and observed off-axis, and if additionally
the field is predominantly normal to the shock plane. In this case we
predict dominance of linear polarization at all frequencies (with the
exception of a comparable circular component at low radio frequencies).
However, this is not an exclusive hallmark of a large scale magnetic
field in the narrow, off-axis jet regime, since a high degree of
linear polarization may result also if the field is entangled over
small length scales, as may have been the case in GRB 021206.

Recently, Matsumiya and Ioka (2003) treated similar propagation effects
in the context of GRB afterglows, neglecting the inhomogeneity of the
emitting slabs due to synchrotron losses. As we demonstrated explicitly in
\S\,\ref{sec:Observational_Consequences_for_Uniform_Strong_Field},
cooling has a decisive effect, most importantly on the values of the
propagation coefficients. This leads to differences between Matsumiya
and Ioka's results and ours (e.g. the higher degree of circular
polarization, by approximately an order of magnitude, that we find
in the transition frequencies 1 GHz and few $\times 10^{13}$ Hz).

Our results suggest that spectro-polarimetric observations in the radio
and IR bands during the early stages of the afterglow may be used as a
unique probe on the structure and strength of the magnetic field. A handle
on the field configuration is of particular importance in studying the
plasma conditions in the reverse shock, since this plasma is essentially
the one outflowing from the compact source, and hence may reveal valuable
information on the processes driving the outflow and on the nature of the
progenitor. This technique, complementary to measurements of the
$\gamma-$ray polarization, may therefore be powerful in placing stringent
constraints on models of the ``inner engine'' and the origin of the
magnetic field, and promote our understanding of the physics of
collisionless shock waves.


\acknowledgements
We thank the anonymous referee for helpful suggestions and instructive
comments.
This research was supported in part by ISF, AEC \& Minerva grants
for E.W., and by grants from NASA (NAG 5-13292) and NSF (AST-0071019,
AST-0204514) for A.L.




\appendix

\section{Propagation, emission and absorption coefficients in
a relativistic plasma}
\label{sec:appendix_A}

\subsection{The propagation coefficients in a relativistic
plasma} \label{sec:Propagation_Coefficients_Appendix}

We briefly review here the derivation of the coefficients $f$ and
$h$ [see Eqs. (\ref{eq:susceptibility_tensor_in_intro_section}) --
(\ref{eq:Transfer_Equation_for_Stokes_Parameters})], associated
with propagation effects, for the case of a plasma with an
isotropic, relativistic electron population. The deviation of the
dielectric tensor
\begin{equation}
\epsilon_{ij}(\omega,{\bf k}) = \delta_{ij} + 4\pi
\kappa_{ij}(\omega,{\bf k})\;
\end{equation}
from its value in vacuum ($\delta_{ij}$) is assumed to be small,
so that the radiation can be regarded as transverse (Sazonov
1969).

To obtain $f$, a quantity which is first order in $B$, we separate
the perturbation to the equilibrium distribution function into two
contributions, one which is independent of the magnetic field, and
one which is linear in the magnetic field. The Maxwell equations
lead to a dispersion relation~:
\begin{equation}\label{eq:wave_equation}
c^2 k^2 / \omega^2 E_i = \epsilon_{ij} E_j \qquad , \qquad i, j =
1,2\;,
\end{equation}
which, when coupled to a linearized Vlasov equation, gives the
non-diagonal elements~:
\begin{equation}\label{eq:epsilon12}
\epsilon_{12} = -\epsilon_{21} = i \frac{\widetilde{\omega}_p^2
\widetilde{\omega}_B}{\omega} \cos \vartheta \int \textrm{d}^3p
\frac{1}{( \omega - k v_{\parallel} )^2} \frac{\textrm{d}
f^{(0)}}{\textrm{d} p} \frac{p_{\perp 1}^2}{\gamma^2 p} \;.
\end{equation}
Here $\widetilde{\omega}_p^2 = 4\pi n_e e^2/m_e$ and
$\widetilde{\omega}_B = e B / m_e c$ are the non-relativistic
(electron) plasma frequency and Larmor frequency, respectively; \,
$\vartheta$ is the angle between $\mathbf{k}$ and $\mathbf{B}$, \,
$\gamma = \sqrt{p^2 + m_e^2 c^2} / m_e c$ is the particle's
Lorentz factor, and $f^{(0)}(p)$ is the unperturbed isotropic
electron distribution function.

Sagiv and Waxman (2002) showed numerically that the value of
$\epsilon_{12}$ is not very sensitive to the exact shape of the
distribution function, and may be approximated by substituting a
mono-energetic electron distribution $f^{(0)}(\gamma) \propto
\delta (\gamma - \gamma_m) $ in equation (\ref{eq:epsilon12}),
with $\gamma_m$ a characteristic electron Lorentz factor
(typically at the peak of the distribution function). This leads
to a simple expression for the propagation coefficient $f$ in the
relativistic regime~:
\begin{equation}\label{eq:f_rel_delta}
f_{rel}(\nu) \simeq \frac{e^3 n_e B \cos \vartheta}{\pi m_e^2 c^2
\nu^2} \frac{\ln \gamma_m}{\gamma_m^2}\;,
\end{equation}

To obtain $h_{rel}$, we started with equation (9) of Sazonov
(1969), who used a kinetic equation approach to obtain analytic
expressions for the elements of the susceptibility tensor
$\kappa_{ij}$. Using the notation of the current paper, this
equation reads~:
\begin{equation}
\kappa_{11} = -\frac{1}{4} \frac{e^2 m_e^2 c^6}{\omega^2} \left(
\frac{\widetilde{\omega}_B \sin \vartheta}{\omega} \right)^2
\int_0^{\infty} \textrm{d}\mathcal{E} \frac{\textrm{d}}
{\textrm{d}\mathcal{E}} \left[ \frac{\textrm{d}n(\mathcal{E})
/\textrm{d}\mathcal{E}}{\mathcal{E}^2} \right] \gamma^4 Y(z)\;,
\end{equation}
with $\mathcal{E} = \gamma m_e c^2$ the particle energy, and
$\textrm{d}n(\mathcal{E})/\textrm{d}\mathcal{E}$ the electron
energy spectrum; $z = (\frac{3}{2} \nu / \nu_{syn})^{2/3}$ is a
normalized frequency, where $\nu_{syn} = 3eB\gamma^2 /4\pi m_e c$
is the characteristic synchrotron frequency for electrons with
Lorentz factor $\gamma$; The function $Y(z)$ is defined as~:
\begin{eqnarray}
Y(z) &=& \frac{z^2}{2} \mathcal{P} \left\{ \int_{-\infty}^{\infty}
\frac{\textrm{d}\nu}{\nu} \exp ( i\nu z + i\nu^3/3 ) \right\} +
\frac{3 z^2}{2} \int_{-\infty}^{\infty} \textrm{d}\nu \,\nu
\exp ( i\nu z + i\nu^3/3 ) \nonumber\\
&+& z^2 \int_0^{\infty} \textrm{d}\nu \,\nu \cos(\nu z+\nu^3/3)\;,
\end{eqnarray}
where $\mathcal{P}$ denotes Cauchy's principal value of the
integral.

Since we are interested in $h$, only the real part of $Y(z)$
matters, and it can be easily shown that in fact only the last
integral contributes. At frequencies $\nu \ll \nu_m$,
$\kappa_{11}$ is dominated by electrons at the peak of the
distribution function. We therefore take $\textrm{d}n(\mathcal{E})
/ \textrm{d}\mathcal{E} \propto \delta(\mathcal{E} - \gamma_m m_e
c^2)$. Expanding $Re [Y(z)]$ at the limit $z \ll 1$, we finally
obtain
\begin{equation}
h_{rel} \simeq \frac{e^4 n_e B^2 \sin^2 \vartheta}{4 \pi^2 m_e^3
c^3 \nu^3} \frac{\gamma_m}{2} \left(
\frac{\nu}{\nu_m(\gamma_m)} \right)^{4/3} \qquad (\nu \ll
\nu_m)\;.
\end{equation}

At frequencies $\nu > \nu_m$ the value of $h_{rel}$ is
sensitive to the shape of the electron distribution function.
Specifically, for a power-law distribution, we simplify equation
(20) of Sazonov (1969) to get the second of equations
(\ref{eq:h_rel})~:
\begin{equation}
h_{rel} \simeq \frac{e^4 n_e B^2 \sin^2 \vartheta}{4 \pi^2 m_e^3
c^3 \nu^3} \left(\frac{2}{p-2} \right) \gamma_m^{-(p-2)} \left(
\frac{\nu}{\nu_m} \right)^{(p-2)/2} \qquad (\nu >
\nu_m)\;.
\end{equation}

\subsection{The emissivity and absorption coefficients
for synchrotron radiation} \label{sec:em_abs_coefs}

For completeness we provide expressions for the emissivity and absorption
coefficient of the synchrotron radiation, using the notation $x
=\nu/\nu_{syn}$ [see, e.g. Pacholczyk (1977), but note the typos in this
source].

\begin{eqnarray}\label{eq:emissivity_coefficients_Pacholczyk}
\left( \begin{array}{c}
\varepsilon_I \\[3 mm]
\varepsilon_Q \\[3 mm]
\varepsilon_V
\end{array} \right) &=&
\frac{\sqrt{3}}{4 \pi} \frac{e^3 B}{m_e c^2} \sin \vartheta
\int_0^{\infty} \frac{\textrm{d}n(\mathcal{E}_e)}
{\textrm{d}\mathcal{E}_e} \times \nonumber\\
&\quad& \times \left( \begin{array}{c}
x\int_x^{\infty} K_{5/3}(z)\textrm{d}z \\[3 mm]
x K_{2/3}(x) \\[3 mm]
\displaystyle \frac{2 \cot \vartheta}{\gamma_e} \left[x K_{1/3}(x)
+ \int_x^{\infty} K_{1/3}(z) \textrm{d}z \right] \end{array}
\right) \textrm{d}\mathcal{E}_e
\end{eqnarray}

\begin{eqnarray}\label{eq:absorption_coefficients_Pacholczyk}
\left( \begin{array}{c}
\kappa \\[3 mm]
q
\end{array} \right) &=&
-\frac{\sqrt{3}}{8 \pi} \frac{e^3 B}{m_e} \sin \vartheta
\frac{1}{\nu^2} \int_0^{\infty} \mathcal{E}_e^2
\frac{\textrm{d}}{\textrm{d}\mathcal{E}_e} \left(
\frac{\textrm{d}n(\mathcal{E}_e)/\textrm{d}\mathcal{E}_e}
{\mathcal{E}_e^2} \right) \times \nonumber\\
&\quad& \times \left( \begin{array}{c}
x\int_x^{\infty} K_{5/3}(z)\textrm{d}z \\[3 mm]
x K_{2/3}(x) \\
\end{array} \right)
\textrm{d}\mathcal{E}_e
\end{eqnarray}

Note that for a synchrotron source $\varepsilon_U = 0, u = 0$ [Pacholczyk,
(1977)], and we may neglect $v$ in comparison to $\kappa$ and $q$.


\end{document}